\documentclass[12pt,preprint,referee]{aastex}

\RequirePackage{lineno}
\usepackage{txfonts}
\usepackage{gensymb}

\newcommand{\be} {\begin{equation}}

\def\pwn {3C~58}
\def\psrj {PSR\,J0205$+$6449}

\newcommand{\fermi}{{\em Fermi}}

\newcommand{\bc}{\begin{center}}
\newcommand{\ec}{\end{center}}
\def\ltsima{$\; \buildrel < \over \sim \;$}
\def\lsim{\lower.5ex\hbox{\ltsima}}
\def\loe{\lower.5ex\hbox{\ltsima}}
\def\gtsima{$\; \buildrel > \over \sim \;$}
\def\gsim{\lower.5ex\hbox{\gtsima}}
\def\goe{\lower.5ex\hbox{\gtsima}}

\def\ltsima{$\; \buildrel < \over \sim \;$}
\def\lsim{\lower.5ex\hbox{\ltsima}}
\def\loe{\lower.5ex\hbox{\ltsima}}
\def\gtsima{$\; \buildrel > \over \sim \;$}
\def\gsim{\lower.5ex\hbox{\gtsima}}
\def\goe{\lower.5ex\hbox{\gtsima}}

\def\ergscm2 {erg\,s$^{-1}$cm$^{-2}$}

\def\cm2 {cm$^{-2}$}

\shortauthors{Fermi-LAT collaboration}
\shorttitle{Fermi-LAT observations of 3C 58}

\begin{document}
\title{Observing and modeling the gamma-ray emission from pulsar/pulsar wind nebula complex \psrj\//3C 58}

\author{Jian Li\altaffilmark{1,2}, Diego F. Torres\altaffilmark{1,2,3}, Ting Ting Lin\altaffilmark{1,2}, Marie-Helene Grondin\altaffilmark{4}, Matthew Kerr\altaffilmark{5}, Marianne Lemoine-Goumard\altaffilmark{4}, Emma de O\~na Wilhelmi\altaffilmark{1,2} }

\altaffiltext{1}{Institute of Space Sciences (ICE, CSIC), Campus UAB, Carrer de Can Magrans, 08193, Barcelona, Spain; Email: jian@ice.csic.es}
\altaffiltext{2}{Institut d'Estudis Espacials de Catalunya (IEEC), 08034 Barcelona, Spain}
\altaffiltext{3}{Instituci\'o Catalana de Recerca i Estudis Avan\c{c}ats (ICREA), E-08010, Barcelona, Spain; Email: dtorres@ice.csic.es}
\altaffiltext{4}{Universit\'e Bordeaux 1, CNRS/IN2P3, Centre d'\'Etudes Nucl\'eaires de Bordeaux Gradignan, 33175 Gradignan, France}
\altaffiltext{5}{Space Science Division, Naval Research Laboratory, Washington, DC 20375, USA}

\begin{abstract}

We present the results of the analysis of 8 years of \emph{Fermi}-LAT data of the pulsar/pulsar wind nebula complex \psrj\//3C 58.
Using a contemporaneous ephemeris, we carried out a detailed analysis of \psrj\ both during its off-peak and on-peak phase intervals.
3C 58 is significantly detected during the off-peak phase interval.
We show that the spectral energy distribution at high energies is the same disregarding the phases considered, and thus that this part of the spectrum is most likely dominated by the nebula radiation. We present results of theoretical models of the nebula and the magnetospheric emission that confirm this interpretation.
Possible high-energy flares from \pwn\/ were searched for, but none was unambiguously identified.

\end{abstract}


\section{Introduction}
\label{intro}

\label{intro}

\pwn\/ is an extended flat-spectrum radio source which was identified as a supernova remnant (SNR) first in radio (Weiler \& Seielstad 1971) and then in optical by H$\alpha$ observations (van den Bergh 1978).
Because of its flat radio spectrum and bright-filled center, \pwn\ was classified as a pulsar wind nebula (PWN, or plerion, Weiler \& Panagia 1978) long before the central pulsar, \psrj\/, was discovered (Murray et al. 2002).
Subsequent radio imaging observations continued to reveal a center-brightened morphology and a compact size of 6'$\times$9' (Green 1986; Reynolds \& Aller 1988; Bietenholz et al. 2001, 2006), which is consistent with the morphology observed in infrared and X-ray bands (Bocchino et al. 2001; Slane et al. 2004, 2008).
Since the jet-torus structure (Slane et al. 2004), filaments and knots (Fesen et al. 2008) observed in \pwn\ resemble those seen in the Crab Nebula, \pwn\ was proposed to be a ``Crab-like" PWN.
The distance of \pwn\ was estimated as 3.2 kpc (Roberts et al. 1993), although there {is} an on-going discussion on this value, given that a recent H I measurement suggests a nearer distance of just 2 kpc (Kothes 2013).
\pwn\/ was proposed to be the remnant of SN 1181, observed by Chinese and Japanese astronomers (Stephenson 1971; Stephenson \& Green 2002).
However, there is controversy on the viability of this connection (see Fesen et al. 2008, Table 3 and Kothes 2013, Table 1 for a discussion).

\psrj\/ is {a pulsar with a rotational period of} 65 ms located in \pwn\/.
It was discovered by \emph{Chandra} in X-rays many years after \pwn\/ was classified as a PWN.
{Timing parameters give} a surface magnetic field estimation of 3.6$\times$10$^{12}$ G, a characteristic age of 5400 yr{, and a very high spin-down luminosity, 2.7$\times$10$^{37}$ erg s$^{-1}$, making it  the third most energetic of the known Galactic pulsars} (Murray et al. 2002).
Camilo et al. (2002) reported the detection of its radio pulsation, which leads the X-ray pulse by $\sim$ 0.10 spin phase.
%
%
%
Because of the high spin-down power of \psrj\/, {it} was expected to shine in gamma rays.
{\psrj\/ was among the first gamma-ray pulsars detected by \emph{Fermi}-LAT (Abdo et al. 2009).}
{Gamma-ray emission from the PWN \pwn\/ was reported in the second \emph{Fermi}-LAT Pulsar Catalog} (Abdo et al. 2013, which we shall refer to as 2PC hereafter).
\pwn\/ was also detected above 10 GeV in the First and Third Catalog of Hard \emph{Fermi}-LAT Sources, suggesting its potential nature as a TeV gamma-ray source (Ackermann et al. 2013, 1FHL;  {Ajello} et al. 2017, 3FHL).
Several imaging atmospheric Cherenkov telescopes have observed \pwn\/ (\emph{Whipple}, Hall et al. 2001; VERITAS, Aliu 2008; MAGIC I, Anderhub et al. 2010) but it was only recently detected by MAGIC II (Aleksi\'c et al. 2014).
PWN models for \pwn\/ have been presented by Bednarek \& Bartosik (2003, 2005), Bucciantini et al. (2011), and Torres et al. (2013); in the latter paper, a comparison among these models is provided. Here, in an effort to understand better the radiation from the complex, we report on further analysis/modeling of \psrj\//3C 58 using more than eight years of \emph{Fermi}-LAT data and the newest response functions.

\section{OBSERVATIONS}
\label{obs}

The \emph{Fermi}-LAT (Atwood et al. 2009) data used for this paper range from 2008 August 4 (MJD 54682) to 2016 September 20 (MJD 57651), covering 97 months and extending the three years' coverage of the 2PC.
\emph{Fermi} Science Tools\footnote{\url{http://fermi.gsfc.nasa.gov/ssc/}},
10-00-05 release was used to analyze the data.
The data selection and analysis method adopted in this paper are similar with those in Li et al. 2017.
We selected events from the ``Pass 8'' event class,
and used ``P8R2 V6 Source'' instrument response functions (IRFs).
All gamma-ray photons within an energy range of 0.1--300 GeV and a circular region of interest (ROI) of 10$\degree$ radius centered on \psrj\/ were considered.
Gamma-ray photons were selected only with a zenith angle $<$ 90$\degree$ to reject contaminating gamma rays from the Earth's limb.

The spectral results presented in this work were calculated by performing a binned maximum likelihood fit (30 logarithmically spaced bins in the 0.1--300 GeV range) using the Science Tool \emph{gtlike}.
A spectral-spatial model was constructed including Galactic and isotropic diffuse emission components (``gll\_iem\_v06.fits", Acero et al. 2016, and ``iso\_P8R2\_CLEAN\_V6\_v06.txt", respectively\footnote{\url{http://fermi.gsfc.nasa.gov/ssc/data/access/lat/BackgroundModels.html}})  as well as known gamma-ray sources within $15\degree$ of the \pwn\/, based on {\textit{Fermi} LAT Third Source Catalog ({3FGL,} Acero et al. 2015)}.
The spectral parameters of the sources within $3\degree $ of our target were left free.
The spectral parameters of other sources included were fixed at the 3FGL values.
In the pulsar phase-resolved analysis, photons within a specific spin phase interval are selected.
To account for that, the prefactor parameters of all sources were scaled to the width of the spin phase interval.
The significance of the sources are evaluated by the Test Statistic (TS).
This statistic is defined as TS=$-2 \ln (L_{max, 0}/L_{max, 1})$, where $L_{max, 0}$ is the maximum likelihood value for a model in which the source studied is removed (the ``null hypothesis"), and $L_{max, 1}$ is the corresponding maximum likelihood value with this source incorporated.
The larger the value of TS, the less likely the null hypothesis is correct (i.e., a significant gamma-ray excess lies on the tested position).
The square root of the TS is approximately equal to the detection significance of a given source.
%
%
The \textit{pointlike} analysis package (Kerr 2011a) was used to produce the TS maps in this paper.
In the analysis, the systematic errors have been estimated by repeating the analysis using modified IRFs that bracket the effective area\footnote{\url{http://fermi.gsfc.nasa.gov/ssc/data/analysis/scitools/Aeff\_Systematics.html}} (Ackermann et al. 2012), and artificially changing the normalization of the Galactic diffuse model  by $\pm$ 6\% (2PC).
{The {latter} dominates the systematic errors.}
{Energy dispersion correction has been adopted in spectral analysis}\footnote{\url{https://fermi.gsfc.nasa.gov/ssc/data/analysis/documentation/Pass8\_edisp\_usage.html}}.
{In this paper, the} first (second) uncertainty shown corresponds to the statistical (systematic) error.

To search for the possible {spatial} extension of {\pwn\/} in the off-peak gamma-ray emission, we followed the method described in Lande et al. (2012).
The source is {modeled} to be spatially extended with a symmetric disk model.
We fitted its position and extension with the \textit{pointlike} analysis package.
The extension significance was defined as {TS$_{ext}$=$-2(\ln L_{point}/L_{disk}$)}, in which $L_{disk}$ and $L_{point}$ were the \textit{gtlike} global likelihood of the extended source hypotheses and the point source, respectively.
A threshold for claiming the source to be spatially extended is set as TS$_{ext}>$16, {which corresponds} to a significance of $\sim$ 4$\sigma$.

\section{Off-peak and on-peak phase selection}

Photons from \psrj\/ within a radius of 0$\fdg$65 and a minimum energy of 200 MeV were selected, which maximized the H-test {statistic} (de Jager et al. 1989; de Jager \& B$\ddot{u}$sching 2010).
Adopting the most updated ephemeris for \psrj\/ (M. Kerr \& D. Smith 2017, private communication), 
we assigned pulsar rotational phases to each gamma-ray photon that passed the selection criteria, using \emph{Tempo2} (Hobbs et al. 2006) with the {\it Fermi} plug-in (Ray et al. 2011).
%

\begin{center}
\begin{figure*}
\centering
\includegraphics[scale=0.8]{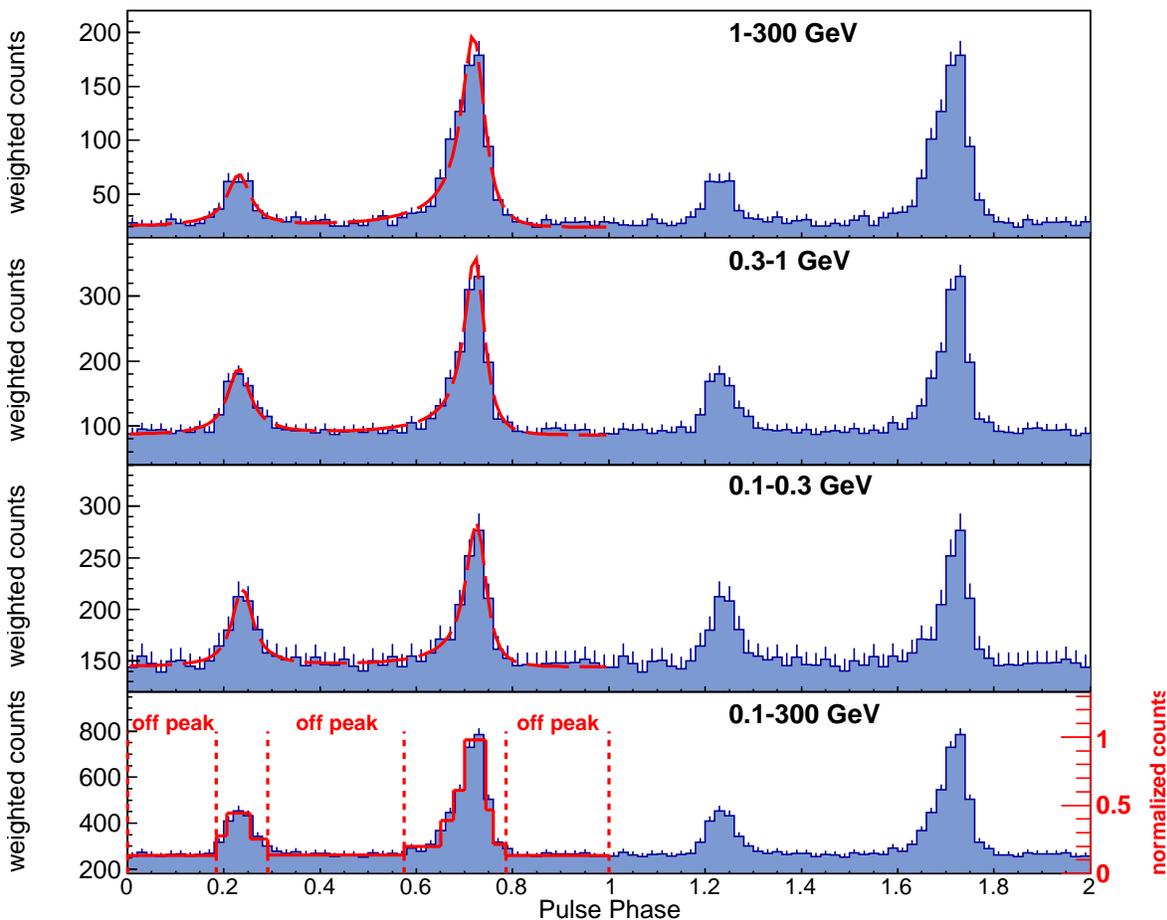}
\caption{Weighted pulse profile of \psrj\/ at different energies.
Two rotational pulse periods are shown, with a resolution of 50 phase bins per period.
The double asymmetric Lorentzian {profiles} plus a constant that we have fitted to the light curves are shown with dashed red lines.
The bottom panel shows the weighted pulse profile above 100 MeV.
The Bayesian block decomposition is represented by red lines in the bottom panel, {and is normalized to the highest block}.
{The red dotted lines define the off-peak intervals.}}
\label{profile}
\end{figure*}
\end{center}

The {light curve} of \psrj\/ was divided into two parts, an off-peak and an on-peak interval.
We began by deconstructing the pulsed light curve into simple Bayesian Blocks using the same algorithm described in the 2PC, by Jackson et al. (2005) and Scargle et al. (2013).
To produce Bayesian Blocks on the light curve, we have extended the data over three rotations, by copying and shifting the observed phases to cover the phase range from $-$1 to 2.
We have defined the final blocks to be between phases 0 and 1.
To increase the statistics and in accordance with the current results, we have adopted a wider interval for the off-peak phases than that used in the 2PC (0.35 of the total).
The off-peak interval in our analysis is then defined to be at $\phi$=0.0$-$0.184, 0.291$-$0.574 and 0.786$-$1.0, yielding 0.681 of the the total revolution.
{We also tested a conservative selection for the off-peak phases, selecting them as $\phi$=0.0$-$0.144 and 0.825$-$1.0, which is defined as the lowest Bayesian block with 10\% reduction on either side (2PC), yielding 0.319 of the total revolution.
It leads to consistent results.}
The on-peak phases are thus located at $\phi$=0.184$-$0.291  and  $\phi$=0.574$-$0.786.
{Figure \ref{profile} shows the pulsar spin light curves, using a photon weighting technique based on the method of Kerr (2011b).}
Additional discussion of Figure \ref{profile} is presented in Section \ref{onpeak}.

\label{results}

  \section{Off-peak analysis: detecting the gamma-ray emission from PWN \pwn\/}

\begin{center}
\begin{figure*}[bt]
\centering
\includegraphics[scale=0.43]{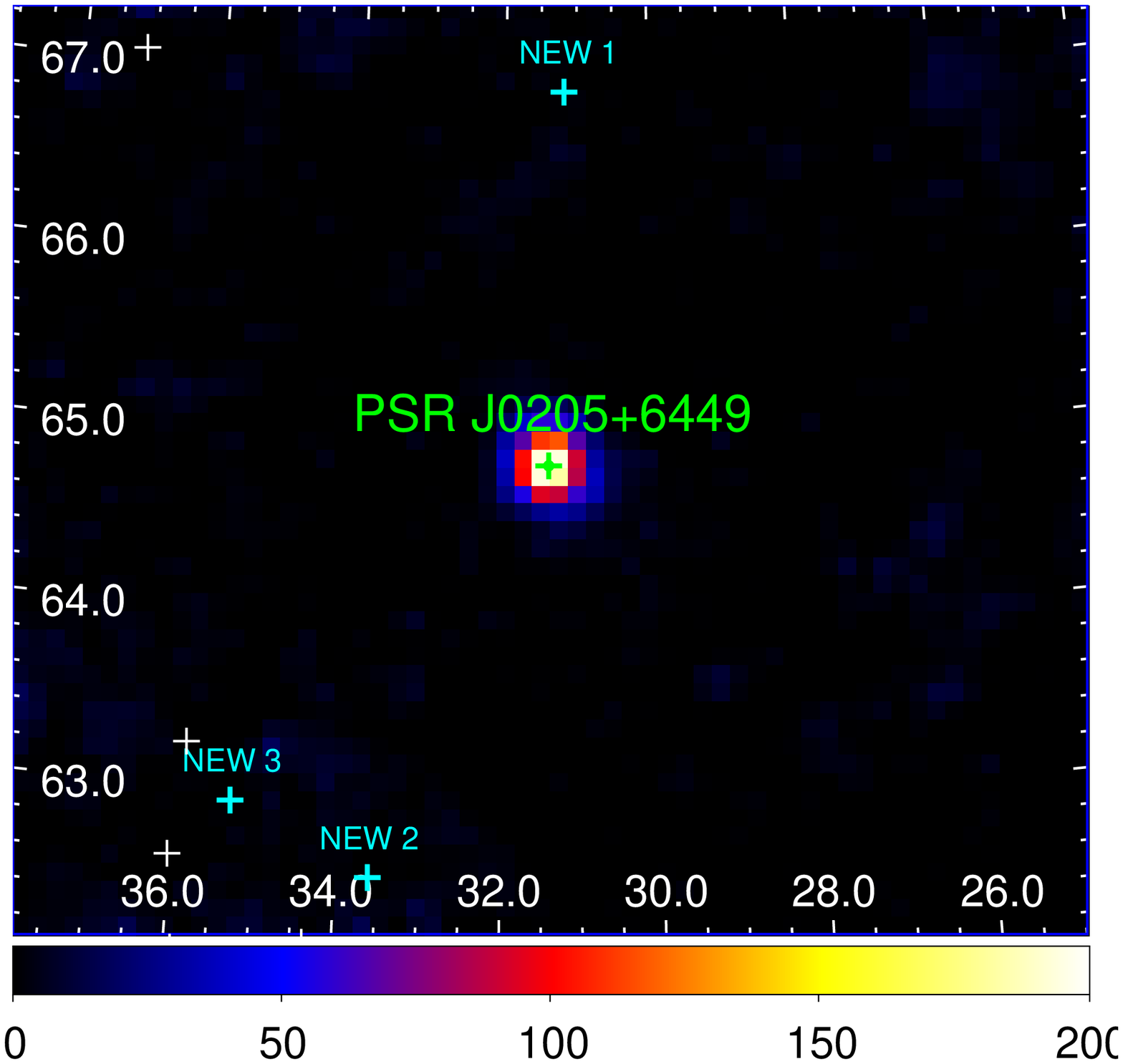}
\includegraphics[scale=0.43]{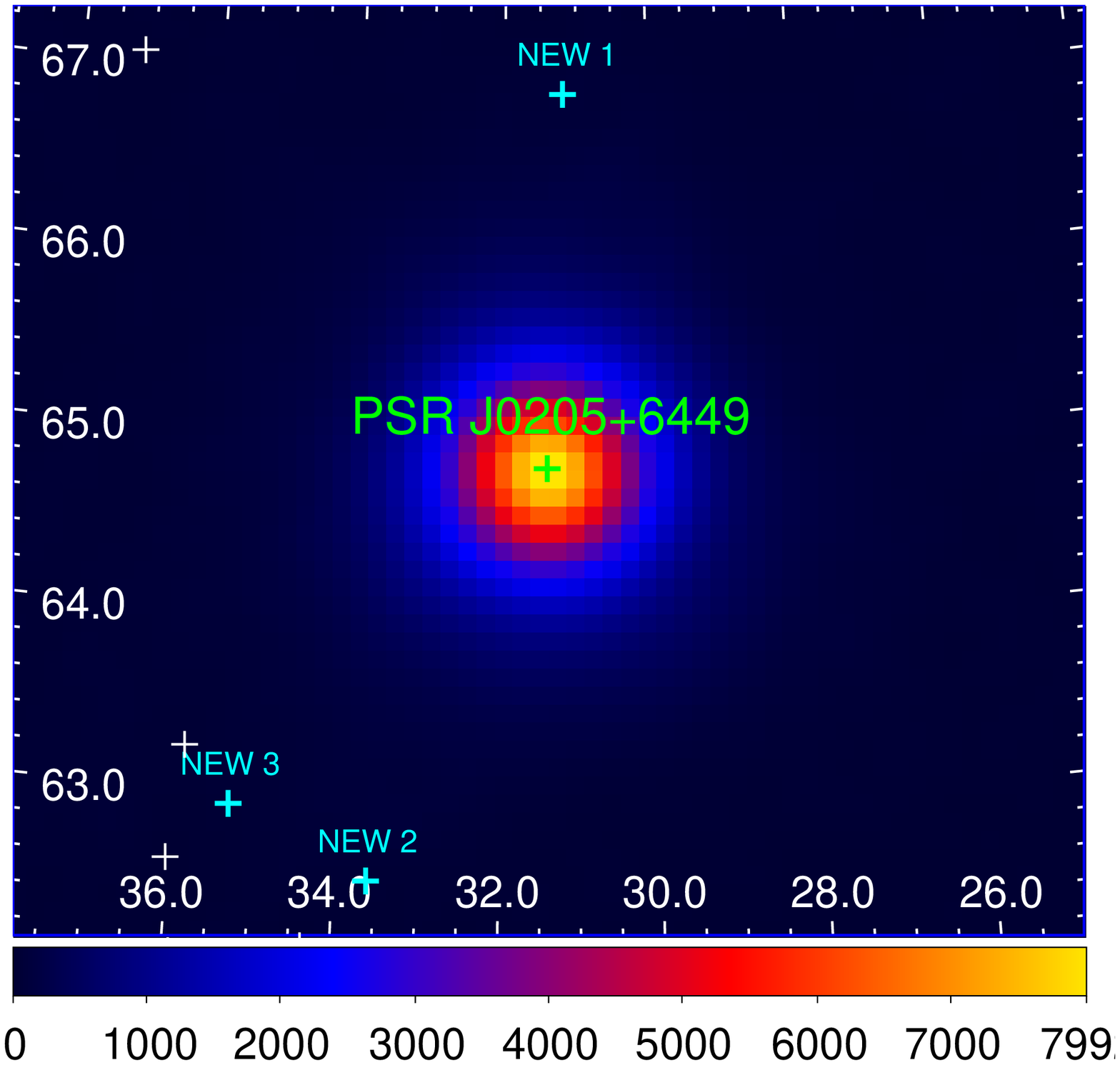}
\caption{TS map (0.1--300 GeV) of the \textit{Fermi}-LAT field surrounding \psrj\ during off-peak (left) and on-peak (right) phases. {The position of \psrj\/} is shown as a green cross while other sources included in the model {from 3FGL} are shown as white crosses. {The new sources added in the analysis are shown with cyan crosses}. The X-and Y-axis are {R.A.} and {decl.} referenced at J2000.}
\label{tsmap}
\end{figure*}
\end{center}

\begin{center}
\begin{figure*}
\centering
\includegraphics[scale=0.41]{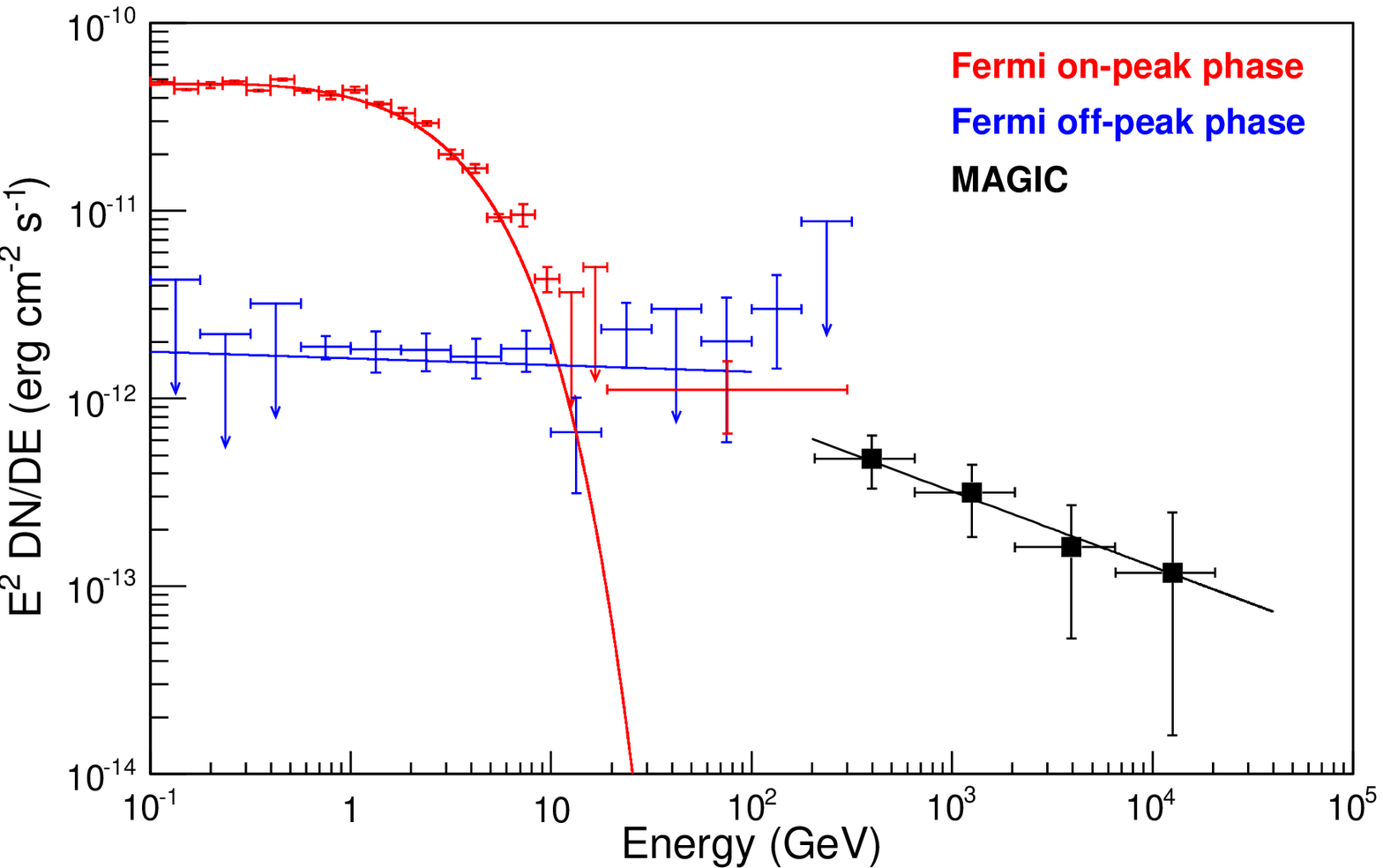}
\includegraphics[scale=0.41]{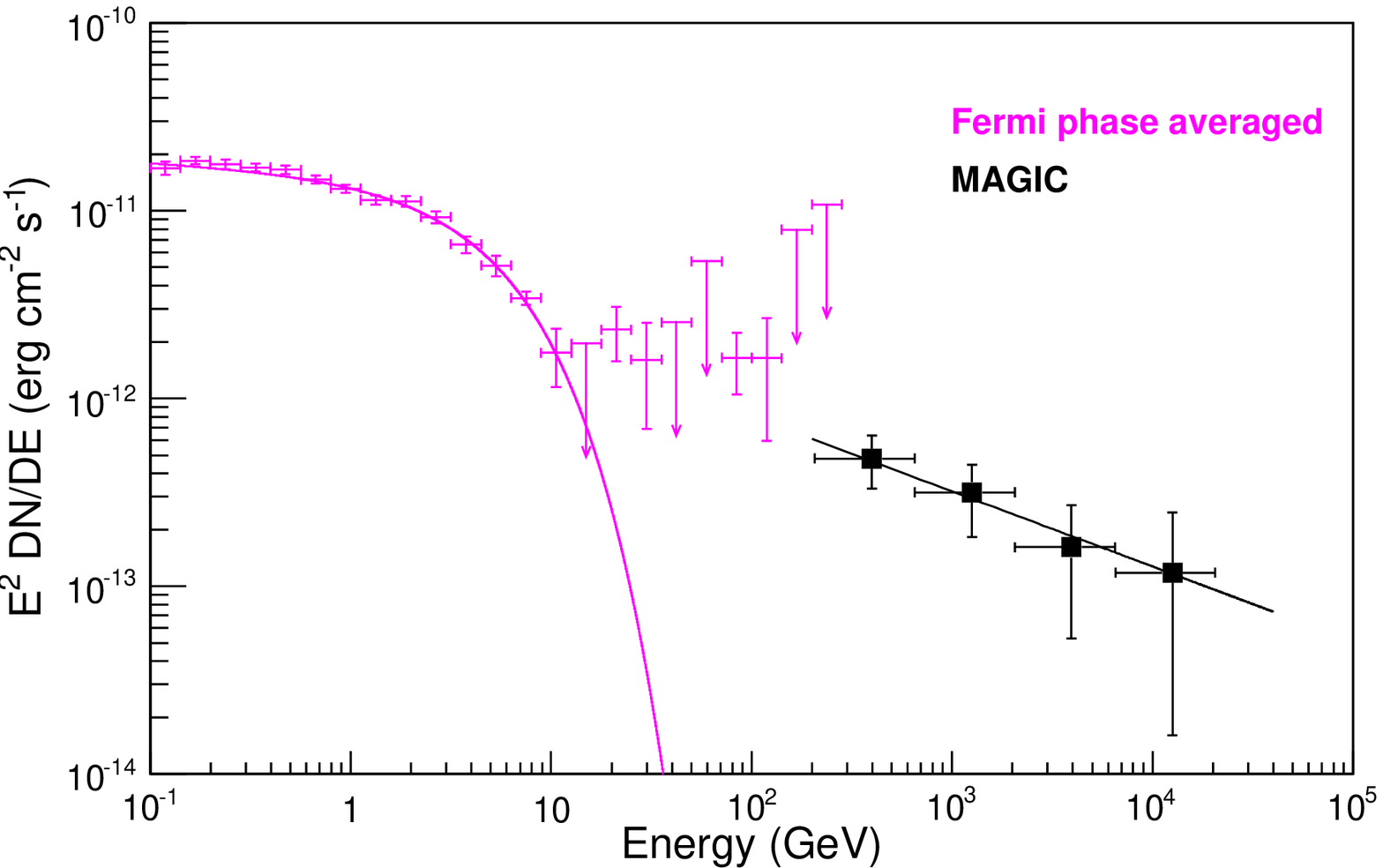}
\caption{Left: {\em Fermi}-LAT spectra of \psrj\/\pwn\ during off-peak (blue) and on-peak (red) phases.
{A 95\% upper limit is calculated if the TS of the SED point is less than 9.}
Maximum likelihood models fitted with \emph{gtlike} are shown with red lines (power law with exponential cutoff) and blue lines (power law).
The MAGIC spectral points and overall fit (Aleksi\'c et al. 2014a) are shown in black for comparison.
Right: {phase-averaged} (purple) {\em Fermi}-LAT {spectrum} of \psrj\/ shown with the maximum likelihood model fitted with \emph{gtlike} (a power law with exponential cutoff).
{The errors shown here are statistical}.
}
\label{psrj_sed}
\end{figure*}
\end{center}

The off-peak emission of {\psrj\//\pwn\/} was previously reported in the 2PC using three years of data with P7V6 IRFs (P7V6 is one of the previous versions of LAT IRFs\footnote{\url{http://fermi.gsfc.nasa.gov/ssc/data/analysis/documentation/Cicerone/Cicerone\_LAT\_IRFs/IRF\_overview.html}}), yielding a TS value of 33.7 and a flux level of (1.75 $\pm$ 0.68) $\times$ 10$^{-11}$ erg~cm$^{-2}$s$^{-1}$ in {the} 0.1--316 GeV {range}. This emission was thought to be the pulsar wind nebula radiation.
3FGL was based on four years of Pass 7 LAT data while in this analysis the dataset is doubled (eight years Pass 8 data).
Thus the region around {\psrj\//\pwn\/} may not be well modelled solely by 3FGL sources.
For a better spectral-spatial modelling of this region, we added three additional point sources (assumed to be described by a simple power law) to the spatial model (Figure \ref{tsmap}), using a similar method as Caliandro et al. (2013).
The best {positions} of the additional sources were determined with \textit{pointlike} as R.A. = 31$\fdg$19$\pm$0.05, decl.= 66$\fdg$91$\pm$0.04 (NEW 1); R.A. = 33$\fdg$59$\pm$0.05, decl.= 62$\fdg$52$\pm$0.04 (NEW 2); R.A. = 35$\fdg$30$\pm$0.04, decl.= 62$\fdg$92$\pm$0.04 (NEW 3).
{Assuming} a power law spectral shape ($dN/dE=N_{0}(E/E_{0})^{-\Gamma}$ cm$^{-2}$ s$^{-1}$ GeV$^{-1}$),
the off-peak analysis of {\psrj\//\pwn\/} {results} in a TS value of 202 and a flux level of (1.24 $\pm$ 0.13 $\pm$ 0.17) $\times$ 10$^{-11}$ erg~cm$^{-2}$s$^{-1}$ in {the} 0.1--300 GeV {energy range}, which is consistent with 2PC results.
The spectral index is estimated as 2.04 $\pm$ 0.07 $\pm$ 0.15 ({Table \ref{psrj_fit}}), which is softer than the value reported in the 2PC (1.61$\pm$0.21) but still within 3 $\sigma$ errors.
Figure \ref{tsmap}, left panel shows the TS map of {\psrj\//\pwn\/} region during off-peak phases.
{In off-peak phases} we also modeled {\psrj\//\pwn\/} with a power law having an exponential cutoff ($dN/dE=N_{0}(E/E_{0})^{-\Gamma}$exp$(-E/E_{0}) $ cm$^{-2}$ s$^{-1}$ GeV$^{-1}$).
The two models are compared using the likelihood ratio test (Mattox et al. 1996).
The $\Delta$TS\footnote{$\Delta$TS=$-2 \ln (L_{PL}/L_{CPL})$, where $L_{CPL}$ and $L_{PL}$ are the maximum likelihood values for power-law models with and without a cut off.} between the two models is {0.03}, which indicates that a cutoff is not significantly preferred. This result is also consistent with the 2PC.
The best-fit spectral parameters and corresponding TS values are listed in Table \ref{psrj_fit}, while the {spectral energy distributions} (SEDs)\footnote{The SEDs are produced by repeating the likelihood analysis in 12 equally spaced logarithmic energy bins, with photon index fixed at 2.04} along with the best-fit power law model are shown in Figure \ref{psrj_sed}, left panel.

The extension of {\psrj\//\pwn\/} during off-peak phases was explored as well in the 2PC analysis, but has not been {favored} over a point-like morphology.
%
%
Here, using \textit{pointlike}, we have fitted an extended disk to the off-peak gamma-ray emission of {\psrj\//\pwn\/}, yielding a $\Delta$TS$_{ext}$ $=$ 0.1; the disk is not {favored} either.
The localization of the off-peak emission determined with \textit{pointlike} is R.A. = 31$\fdg$40, decl.= 64$\fdg$83, with a 95\% confidence error circle of radius 0$\fdg$025, which is consistent with {\psrj\//\pwn\/}.
Considering the flat spectrum and the absence of a spectral cutoff, it is natural to propose that the off-peak gamma-ray emission of {\psrj\//\pwn\/} is dominated by the PWN \pwn\/, {though a weak magnetospheric component cannot be completely ruled out at low energies}.
{\pwn\ was also detected in the 1FHL and the 3FHL catalogs, and the reported spectra are consistent with our results.}
The detected morphology of \pwn\/ being point-like in 0.1--300 GeV is not unexpected.
The arcmin-sized extension in radio and X-rays (Bietenholz et al. 2001; Bocchino et al. 2001) is smaller than the \emph{Fermi}-LAT PSF (e.g. 0$\fdg$1 at 10 GeV ).
%

%

\section{On-peak analysis: studying the magnetospheric emission from \psrj\/}
\label{onpeak}

\begin{table*}{}
\centering
\scriptsize
\caption{{\em Fermi}-LAT spectral parameters of \psrj\ during off-peak and on-peak phases.}
\begin{tabular}{cccccc}
\\
\\
\hline\hline
Phase Interval&Spectral Index& Cutoff Energy  &TS&Flux, 0.1--300 GeV  \\
                &                          &   (GeV)          &       &10$^{-11}$ erg~cm$^{-2}$s$^{-1}$\\

\\
\hline\hline                 
\\
phase averaged         &  2.06 $\pm$ 0.04 $\pm$ 0.14   &  5.05 $\pm$ 0.07 $\pm$ 1.22 & 4213  & 5.53 $\pm$ 0.13 $\pm$ 0.46 \\

off-peak         &  2.04 $\pm$ 0.07 $\pm$ 0.15      &  -   &202  &1.24 $\pm$ 0.13 $\pm$ 0.17 \\

on-peak                       &  1.93 $\pm$ 0.04 $\pm$ 0.08   &  2.87 $\pm$ 0.03 $\pm$ 0.46 & 7718  & 15.31 $\pm$ 0.27 $\pm$ 0.88 \\
\\
\\
\hline\hline                 
\\

\tablecomments{The first (second) uncertainties correspond to statistical (systematic) errors.}
\label{psrj_fit}
\end{tabular}
\end{table*}

We have considered a power law with an exponential cutoff for modeling the on-peak emission of \psrj.
The best-fit parameters are shown in Table {\ref{psrj_fit}}.
The right panel of
Figure \ref{tsmap} shows the on-peak TS map of the \psrj\ region.
Its SED is shown in Figure \ref{psrj_sed}.
At high energies, the SED is consistent with that derived for the off-peak, which indicates that the PWN \pwn\ dominates the flux.
Alternative spectral shapes, like a power law with a sub-exponential cutoff ($dN/dE=N_{0}(E/E_{0})^{-\Gamma}\exp(-E/E_{0})^{b} $ cm$^{-2}$ s$^{-1}$ GeV$^{-1}$, leaving the exponential index $b$ free) yield a $\Delta$TS$= 5$ and are thus not preferred.

We have also modeled the {phase-averaged} emission of \psrj\/ with a power law having an exponential cutoff.
The best-fit parameters are shown in Table {\ref{psrj_fit}} and the SED is shown in Figure \ref{psrj_sed}, right panel.
Adopting the best-fit spectral model derived, we calculated the probabilities for photons to come from \psrj\/ within a radius of 3$\degree$, using the tool \textit{gtsrcprob}, and produced a weighted pulsed light curve based on these photons {(Kerr 2011b)}.
The bottom panel of Figure \ref{profile} shows the folded, pulsed light curve above 100 MeV.
The remaining panels of the same figure show the light curve in narrower energy bands.
The light curve shows two distinct peaks, which is consistent with the profile reported by Abdo et al. (2009) and the 2PC.
To locate the two peaks, we fitted the light curve with two asymmetric Lorentzian functions plus a constant (Figure \ref{profile}).
The fitted constant accounts for the background, which as we have just shown, is dominated by the PWN \pwn\/.
The first (P1) and second (P2) peaks are at 0.234$\pm$0.003  and 0.719$\pm$0.001, respectively.
The separation between the two peaks is 0.485$\pm$0.003, which is consistent with Abdo et al. (2009).
The phase reference used in this paper is different from Abdo et al. (2009).
{By shifting $\sim$ 0.149 spin phase, P2 would be aligned with the profile in Abdo et al. (2009).
In that case, the gamma-ray profile we observed is in good alignment with the X-ray profile but is offset from the radio pulse by $\sim$ 0.085. }

\begin{center}
\begin{figure*}[hbt]
\centering
\includegraphics[scale=0.41]{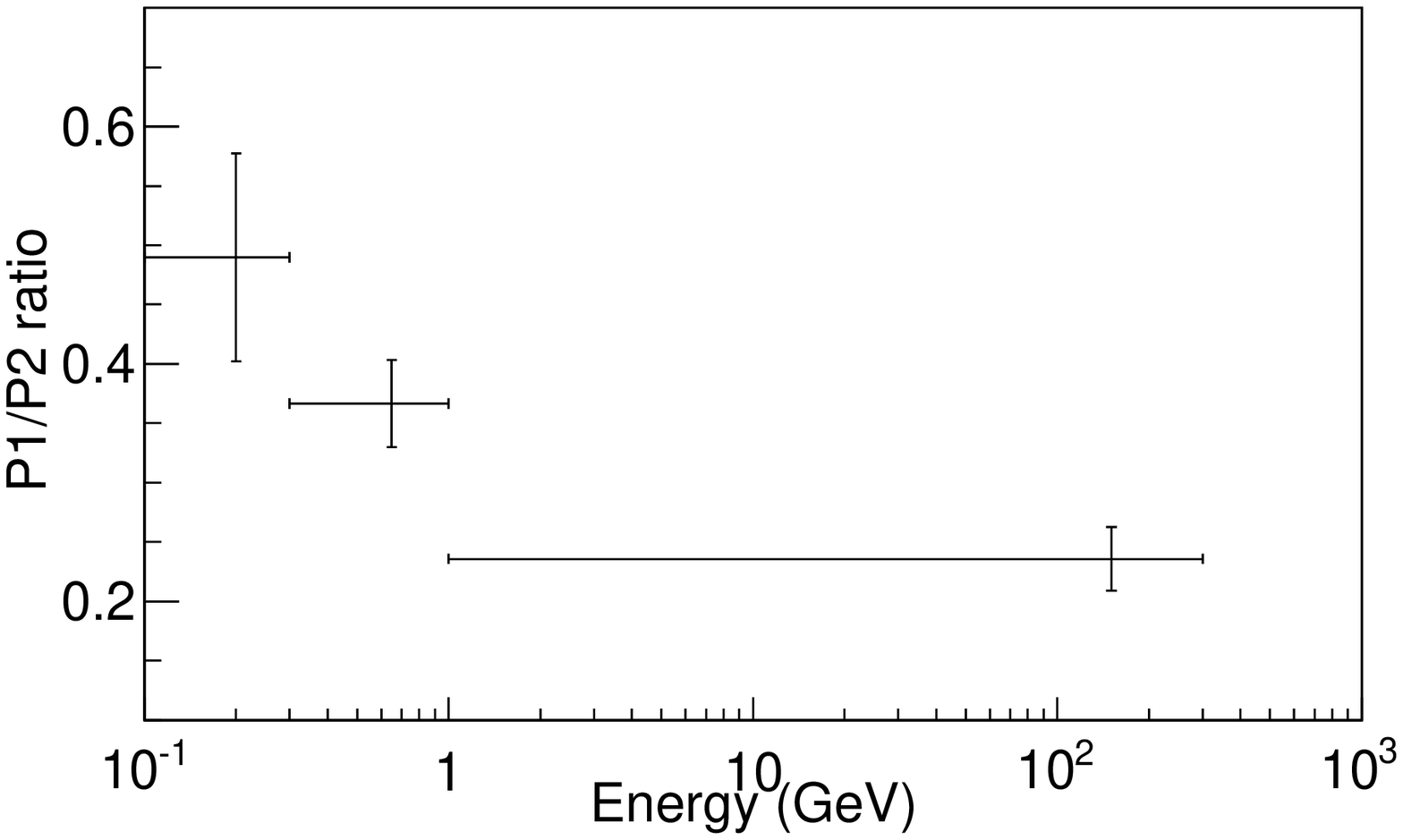}
\includegraphics[scale=0.41]{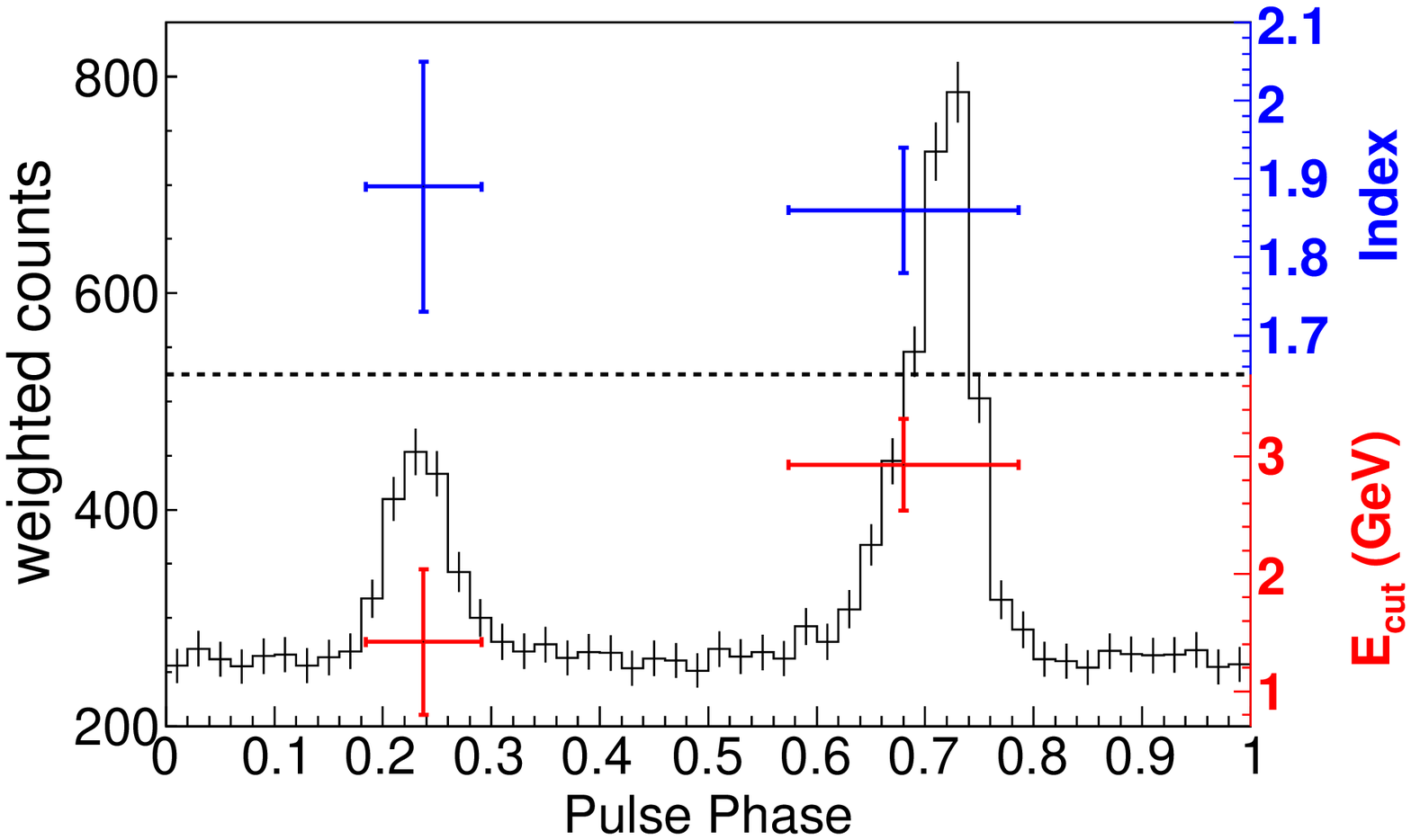}
\caption{Left: Energy evolution of the P1/P2 ratio.
The energy bins are the same as in Figure \ref{profile}.
Right: Spectral parameters of \psrj\/ during the two on-peak intervals.
The histogram shows the weighted phaseogram of \psrj\/ for energies between 0.1 and 300 GeV (similar to Figure \ref{profile}, bottom panel).
The red and blue points correspond to the cutoff energy and the spectral index of the power law with exponential cutoff model, respectively.
{Systematic errors have also been considered. }}
\label{resolved}
\end{figure*}
\end{center}

The strength of P1 and P2 is calculated as the sum of the weighted counts during corresponding on-peak phase ($\phi_{1}$=0.184$-$0.291 and $\phi_{2}$=0.574$-$0.786, respectively) minus the background.
The relative strength of P1 and P2 decreases significantly from low to high energies (Figure \ref{resolved}, left panel), {as} first reported by Abdo et al. (2009).
A
similar trend was observed in Vela, Crab, Geminga, B1951+32, and J0007+7303 pulsars (Thompson 2001; Kanbach 1999; Aleksi\'c et al. 2014b; Li et al. 2016), which {shows} a spectral energy dependence of the gamma-ray light curve.
We carried out spectral analysis for the two peaks.
In the corresponding on-peak phase for the two peaks ($\phi_{1}$=0.184$-$0.291 and $\phi_{2}$=0.574$-$0.786), the pulsar spectrum is modeled as a power law with an exponential cutoff.
The right panel of Figure \ref{resolved} shows the spectral parameters of \psrj\/ in the two on-peak phases.
The lower cut-off energy of P1 when compared with that obtained for P2 (Figure \ref{resolved}, right panel) explains the energy evolution of the P1/P2 ratio.

In order to search for Crab-like flares (Abdo et al. 2013; Buehler et al. 2012) from \pwn\/, we produced light curves with a 30-days time bin in {the} 0.1--1 GeV {and} 1--10 GeV {bands} during {the} off-peak phase and in 10--300 GeV during all spin {phases}. 
The spectral index is fixed at the best-fit value along the off-peak phases, as listed in Table \ref{psrj_fit}.
%
%
{All the light curve data points are below the detection threshold of TS=25; therefore we find no evidence of flaring on this time scale.}
%

\section{Summary and discussion}
\label{discussion}

Using 8.5 years of \emph{Fermi}-LAT data and a contemporaneous ephemeris, we carried out a detailed analysis of \psrj\
both during its off-peak and on-peak phase intervals.

\begin{center}
\begin{figure*}
\centering
\includegraphics[scale=0.42, angle=-90]{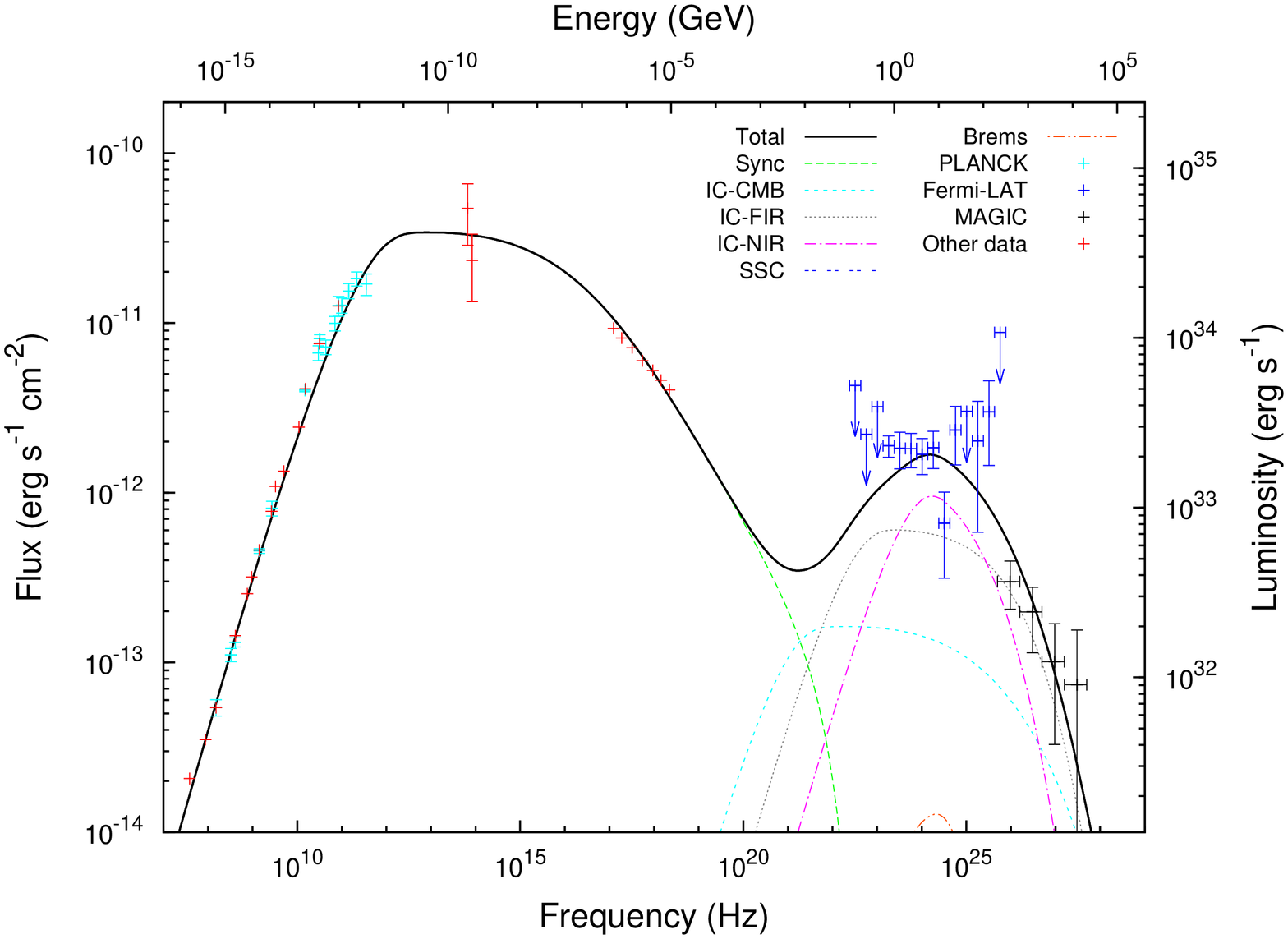}\\
\includegraphics[scale=0.633]{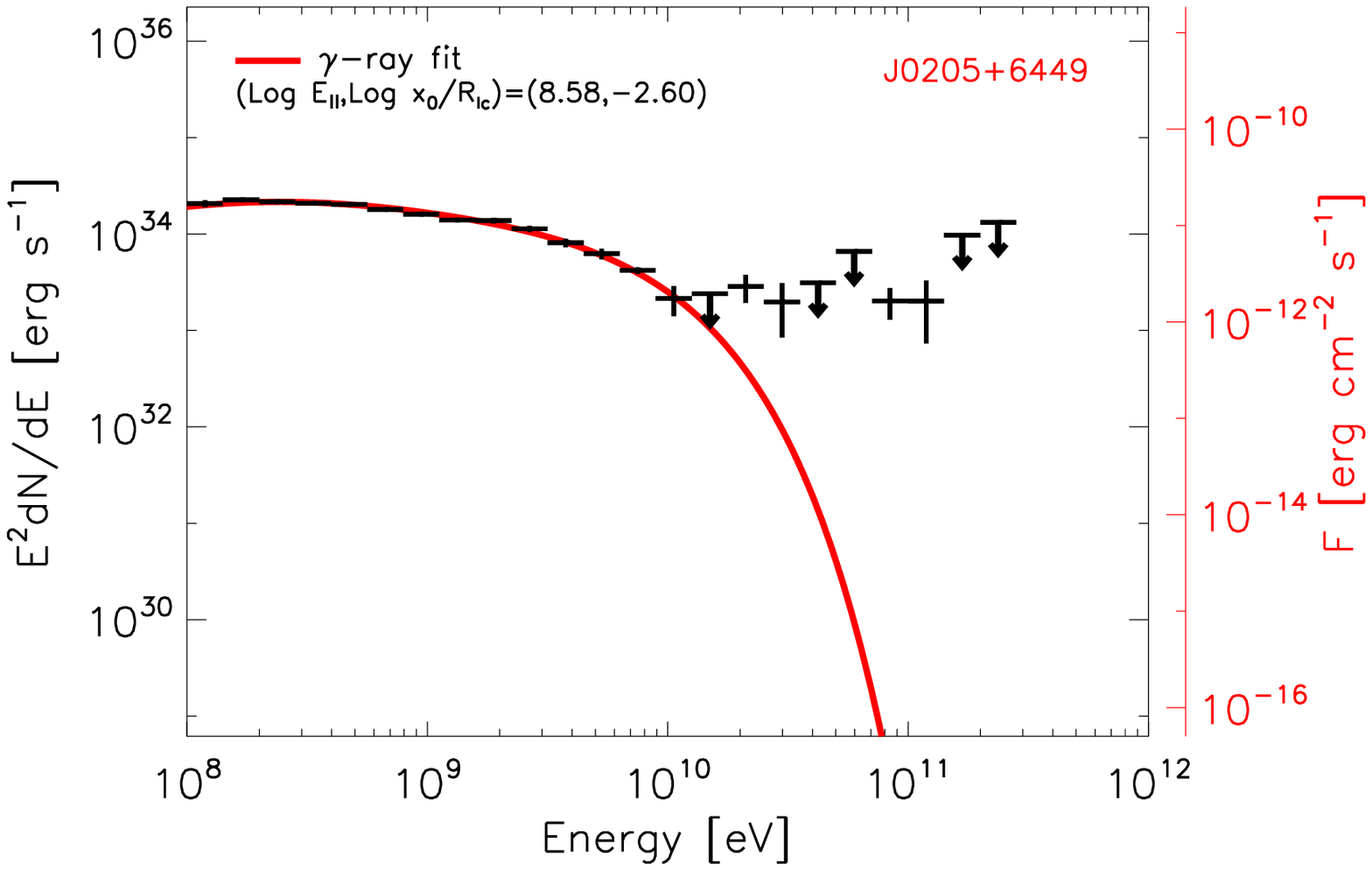}
\caption{Top: {Data for 3C~58 (including the MAGIC and derived off-peak data) compared with a time-dependent, energy-conserving PWN model prediction. The \emph{Planck} data (cyan) are taken from Arnaud et al. (2016) while other data are taken from the compilation used by
Torres et al. (2013).}  Bottom: phase-averaged SED compared with a synchro-curvature model. See text for details.
}
\label{theor}
\end{figure*}
\end{center}

During the off-peak phases, {\psrj\//\pwn\/} is significantly detected, having a TS value of 202.
Its spectrum can be modeled by a simple power law.
No extension is detected.
The flat spectrum and the non-detection of a spectral cutoff argue for a PWN origin of the off-peak gamma-ray emission of {\psrj\//\pwn\/}.
%
%
%
The top panel of Figure \ref{theor} shows a theoretical model of the nebula, based on
a time-dependent integration of the dynamical evolution of both the nebula and the supernova remnant, the radiation of particles, and the particle population.
For details on the model see Martin et al. (2016) and the appendix in Torres (2017).
{The spin-down power, the particle injection, the energy losses, and the
magnetic field all depend on time, and their dependence are accounted for in the model.
The particle content of the nebula is obtained from the balance of energy losses, injection, and escape.
We include losses by synchrotron, inverse-Compton (Klein Nishina inverse Compton with the cosmic-microwave background as well as with IR/optical photon fields), self-synchrotron Compton, and bremsstrahlung, devoid of any radiative approximations, and compute the radiation produced by each process.
The model also considers the dynamical influence of the reverse shock travelling backwards towards the pulsar,
and compressing the nebula.}
However, given that the pulsar is young, this effect is not found to be relevant and results would be very similar if neglected altogether: the nebula is {freely} expanding.
We find a good agreement with data considering a nebula at 3.2 kpc (the same distance we use below for the computation of the pulsar magnetospheric power) and an age of 2500 years.
The fitting model features a broken {power law} for injected electrons at the termination shock, with a low (high) energy index of 1.1 and 2.94, and an energy break at  Lorentz factor $9 \times 10^4$.
These values of the parameters (as well as the magnetic {energy} fraction) are in agreement with our earlier analysis (Torres, Cillis, \& Martin 2013).
The magnetic fraction (the fraction of spin-down energy that goes into the magnetic field) is 0.2; thus 3C~58 is a particle-dominated nebula; {Nevertheless} the magnetic field has one {of the} highest energy reservoirs when compared with all other nebulae of similar age, perhaps with exception of CTA 1 (see Torres et al. 2014).
Using the current data, we have seen that there is a degeneracy regarding which inverse Compton contribution dominates at high energies.
We have explored about a thousand models varying the energy densities and temperatures for the NIR and FIR photon backgrounds and the best fitting one has similar contributions of both.
Thus, we can find models where one or the other dominates without changing the overall fit significantly (e.g., within a factor of 1.3 of the miminum $\chi^2$).
Further data would be needed to distinguish among these possibilities; in particular, \pwn\ will be a bright source for {the Cerenkov Telescope Array}. {Observations} with this facility will help {determine} the peak and the fall-off of the gamma-ray emission, distinguishing between NIR- or FIR-dominated scenarios.

For the on-peak interval, \psrj\ can be modeled by a power law with an exponential cutoff.
We explored the existence of a sub-exponential cutoff, but no {improvement was found}.
\psrj\/ shows a two-peak pulse profile.
The ratio of P1 and P2 decreases significantly with energy (Figure \ref{resolved}, left panel).
This is consistent with the cut-off energy of P1 being lower than that obtained for P2 (Figure \ref{resolved}, left panel).

The most common interpretation of the magnetospheric radiation for this and all other gamma-ray pulsars
is that it originates in synchro-curvature radiation in a high-altitude gap (alternatively see, e.g., Cerutti et al. 2015).
To consider further this statement we have applied the model discussed in detail in Vigano \& Torres (2015) and Vigano, Torres, \& Martin (2015).
In this model, the detectable radiation coming from the magnetosphere can be estimated by integrating the single-particle synchro-curvature spectrum along the travelled distance in the gap, {convolved} with an effective particle distribution, $dN/dx$, where $x$ is the distance along the magnetic field line (see Vigano et al. 2015 for more details on the computation of the synchro-curvature power).
In this model, two parameters, the electric field $E_{||}$ accelerating particles, and the degree of uniformity of the particle's distribution along the trajectory, $x_0/R_{\rm lc}$, define the spectrum completely.
The normalization $N_0$ moves the spectrum up and down in luminosity, without modifying its shape.
We have applied this model to the phase-averaged SED (i.e., \fermi-LAT data only and a fixed magnetic gradient $b = 2.5$, as in Vigano, Torres, \& Martin 2015) and find a good match for the values shown in Figure \ref{theor}, bottom panel.
The fitting results are consistent with those in Vigano et al. 2015.
Our current analysis gives $\log N_{0}=30.90_{-0.30}^{+0.21}$ compared to the value of $\log N_{0}=29.93_{-0.33}^{+0.67}$ in Vigano et al. 2015.
Whereas applying the model to the on-peak phase spectrum only would lead to comparable results (given that the spectral index of both the phase averaged and on-peak
spectra are
the same within errors, and the cutoff energy is only slightly different), the phase-averaged fit allows a comparison with other gamma-ray pulsars for which
only averaged spectral results are currently available. \psrj's
fit parameters (that result well in agreement with those found using earlier data -- see Table 2 of Vigano, Torres, \& Martin 2015)
confirm the reported trend relating $E_{||}$ to $x_0/R_{\rm lc}$ (see the first panels in Figure 2 of Vigano, Torres, \& Martin 2015), emphasizing that synchro-curvature
dominated radiation is likely behind the pulsations.
We note that PSR J0205+6449 is part of the soft gamma-ray pulsar catalog and it has also been detected in X-rays (Kuiper \& Hermsen 2015).
For a complete model of the pulsed spectrum considering also the X-ray
data, and further discussion, see Torres et al. 2018.

\acknowledgments

The \textit{Fermi} LAT Collaboration acknowledges generous ongoing support
from a number of agencies and institutes that have supported both the
development and the operation of the LAT as well as scientific data analysis.
These include the National Aeronautics and Space Administration and the
Department of Energy in the United States, the Commissariat \`a l'Energie Atomique
and the Centre National de la Recherche Scientifique / Institut National de Physique
Nucl\'eaire et de Physique des Particules in France, the Agenzia Spaziale Italiana
and the Istituto Nazionale di Fisica Nucleare in Italy, the Ministry of Education,
Culture, Sports, Science and Technology (MEXT), High Energy Accelerator Research
Organization (KEK) and Japan Aerospace Exploration Agency (JAXA) in Japan, and
the K.~A.~Wallenberg Foundation, the Swedish Research Council and the
Swedish National Space Board in Sweden. Additional support for science analysis during the operations phase is gratefully acknowledged from the Istituto Nazionale di Astrofisica in Italy and the Centre National d'\'Etudes Spatiales in France. {This work performed in part under DOE Contract DE-AC02-76SF00515.}

We acknowledge the assistance from Dr. M. Kerr and Dr. D. Smith with the gamma-ray ephemeris for \psrj\/ and Dr. P. Saz Parkinson for discussions.
We acknowledge the support from the grants AYA2015-71042-P, SGR 2014-1073 and the National Natural Science Foundation of
China via NSFC-11473027, NSFC-11503078, NSFC-11133002, NSFC-11103020, NSFC-11673013, NSFC-11733009, XTP project XDA 04060604 and the Strategic Priority Research Program ``The Emergence of Cosmological Structures" of the Chinese Academy of Sciences, Grant No. XDB09000000, as well as the CERCA Programme of the Generalitat de Catalunya. Work at NRL is supported by NASA.


\begin{thebibliography}{99}

\bibitem[Abdo A., et al. (2010)]{abdo2009} Abdo, A. A., Ackermann, M., Ajello, M. et al. 2009, ApJL, 699, 102

\bibitem[Abdo A., et al. (2010)]{abdo2009} Abdo, A. A., Ajello, M., Allafort, A., et al. 2013, ApJS, 208, 17 (2PC)
\bibitem[Abdo A., et al. (2010)]{abdo2009} Abdo, A. A., Ackermann, M., Ajello, M. et al. 2013, Science, 331, 739.

\bibitem[Abdo A., et al. (2010)]{abdo2009} Ackermann, M., Ajello, M., Allafort, A. et al. 2013, ApJS, 209, 34 (1FHL)
\bibitem[Abdo A., et al. (2010)]{abdo2009} Ajello, M., Atwood, W. B, Baldini, L. et al. 2017, ApJS, 232, 18 (3FHL)
\bibitem[Abdo A., et al. (2010)]{abdo2009} Aleksi\'c, J.,  Ansoldi, S., A., Antonelli, L., A., et al. 2014, A\&A, 567, 8
\bibitem[Abdo A., et al. (2010)]{abdo2009} Aliu, E. 2008, in AIP Conf. Ser. 1085, eds. F. A. Aharonian, W. Hofmann, \& F. Rieger, 324
\bibitem[Abdo A., et al. (2010)]{abdo2009} Arnaud, M., Ashdown, M., Atrio-Barandela, F. et al. 2016, A\&A, 586, 134
\bibitem[Abdo A., et al. (2010)]{abdo2009} Anderhub, H., Antonelli, L. A., Antoranz, P., et al. 2010, ApJ, 710, 828


\bibitem[Abdo A., et al. (2010)]{abdo2009} Atwood, W. B., Abdo, A. A., Ackermann, M., et al. 2009, ApJ, 697, 1071


\bibitem[Abdo A., et al. (2010)]{abdo2009} Bednarek, W., \& Bartosik, M. 2003, A\&A, 405, 689
\bibitem[Abdo A., et al. (2010)]{abdo2009} Bednarek, W., \& Bartosik, M. 2005, JPhG, 31, 1465



\bibitem[Abdo A., et al. (2010)]{abdo2009} Bietenholz, M. F., Kassim, N. E., Weiler, K. W., 2001, ApJ, 560, 772
\bibitem[Abdo A., et al. (2010)]{abdo2009} Bietenholz, M. F. 2006, ApJ, 645, 1180

\bibitem[Abdo A., et al. (2010)]{abdo2009} Bocchino, F., Warwick, R. S., Marty, P. et al. 2001, A\&A, 369, 1078

\bibitem[Abdo A., et al. (2010)]{abdo2009} Bucciantini, N., Arons, J., \& Amato, E. 2011, MNRAS, 410, 381

\bibitem[Abdo A., et al. (2010)]{abdo2009} Buehler, R., Scargle, J. D., Blandford, R. D. et al. 2012, ApJ, 749, 26B
\bibitem[Abdo A., et al. (2010)]{abdo2009} Caliandro, G. A., Hill, A. B., Torres, D. F., et al. 2013, MNRAS, 436, 740
\bibitem[Abdo A., et al. (2010)]{abdo2009} Camilo, F., Stairs, I. H., Lorimer, D. R. et al. 2002, ApJ, 571, L41

\bibitem[Abdo A., et al. (2010)]{abdo2009} Cerutti, B. Philippov, A. A.; Spitkovsky, A., MNRAS, 457, 2401
\bibitem[Abdo A., et al. (2009)]{abdo2009} de Jager, O. C., \& B$\ddot{u}$sching, I. 2010, A\&A, 517, L9
\bibitem[Abdo A., et al. (2009)]{abdo2009} de Jager, O. C., Raubenheimer, B. C., \& Swanepoel, J. W. H. 1989, A\&A, 221, 180
\bibitem[Abdo A. et al.\ (2010)]{abdo2009}Fesen, R. A., Rudie, G., Hurford, A., \& Soto, A. 2008, ApJS, 174, 379


\bibitem[Abdo A. et al.\ (2010)]{abdo2009}Green, D. A. 1986, MNRAS, 218, 533


\bibitem[Abdo A., et al. (2010)]{abdo2009} Hall, T. A., Wakely, S. P., \& VERITAS Collaboration. 2001, in Int. Cosmic Ray Conf., 6, 2485

\bibitem[Abdo A., et al. (2010)]{abdo2009} Hobbs, G., Edwards, R., \& Manchester, R. 2006, Chin. J. Astron. Astrophys. Suppl., 6, 189

\bibitem[Abdo A., et al. (2010)]{abdo2009} Jackson, B., Scargle, J. D., Barnes, D., et al. 2005, ISPL, 12, 105
\bibitem[Abdo A., et al. (2009)]{abdo2009} Kanbach, G., 1999, ApL \& C, 38, 17
\bibitem[Abdo A. et al.\ (2009)]{abdo2009} Kerr, M. 2011a, PhD thesis, Univ.\ Washington, arXiv: 1101.6072
\bibitem[Abdo A. et al.\ (2009)]{abdo2009} Kerr, M. 2011b, ApJ, 732, 38

\bibitem[Abdo A. et al.\ (2009)]{abdo2009}  Kothes, R. 2013, A\&A, 560, 18
\bibitem[Abdo A., et al. (2010)]{abdo2009}  Kuiper, L., \& Hermsen, W. 2015, MNRAS, 449, 3827

\bibitem[Abdo A. et al. (2009)]{abdo2009} Lande, J., Ackermann, M., Allafort, A., et al., 2012, ApJ, 756, 5
\bibitem[Abdo A. et al. (2009)]{abdo2009} Li, J., Torres, D., de On$\tilde{a}$ Wilhelmi, E., et al. 2016, ApJ, 831, 19
\bibitem[Abdo A. et al. (2009)]{abdo2009} Li, J., Torres, D., Cheng, K.-S.; de On$\tilde{a}$ Wilhelmi, E., et al. 2017, ApJ, 846, 169
\bibitem[Abdo A., et al. (2010)]{abdo2009} Martin, J., Torres, D. F. \& Pedaletti, G., 2016, MNRAS, 459, 3868

\bibitem[Abdo A. et al. (2009)]{abdo2009} Murray, S. S., Slane, P. O., Seward, F. D., Ransom, S. M., \& Gaensler, B. M., 2002, ApJ, 568, 226



\bibitem[Abdo A. et al.\ (2010)]{abdo2009} Ray, P.~S., Kerr, M., Parent, D., et al. 2011, ApJS, 194, 17

\bibitem[Abdo A. et al.\ (2010)]{abdo2009} Reynolds, S. P., \& Aller, H. D. 1988, ApJ, 327, 845
\bibitem[Abdo A. et al.\ (2010)]{abdo2009}Roberts, D. A., Goss, W. M., Kalberla, P. M. W., Herbstmeier, U., \& Schwarz, U. J. 1993, A\&A, 274, 427


\bibitem[Abdo A., et al. (2010)]{abdo2009} Scargle, J. D., Norris, J. P., Jackson, B., \& Chiang, J. 2013, ApJ, 764, 167
\bibitem[Abdo A., et al. (2010)]{abdo2009} Slane, P., Helfand, D. J., Reynolds, S. P., et al. 2008, ApJL, 676, 33
\bibitem[Abdo A., et al. (2010)]{abdo2009} Slane, P., Helfand, D. J., van der Swaluw, E., \& Murray, S. S. 2004, ApJ, 616, 403


\bibitem[Abdo A., et al. (2010)]{abdo2009} Stephenson, F. R. 1971, QJRAS, 12, 10
\bibitem[Abdo A., et al. (2010)]{abdo2009} Stephenson, F. R., \& Green, D. A. 2002, Historical Supernovae and Their Remnants (New York: Oxford Univ. Press)


\bibitem[Abdo A., et al. (2010)]{abdo2009} Thompson, D. J. 2001, in AIP Conf. Proc. 558, High Energy Gamma Ray Astronomy, ed. F. A. Aharonian \& H. J. Volk (Melville, NY: AIP), 103
\bibitem[Abdo A., et al. (2010)]{abdo2009} Torres, D. F., Cillis A. N. \& Mart\'in J. 2013, ApJ, 763, 4
\bibitem[Abdo A., et al. (2010)]{abdo2009} Torres, D. F. 2017, ApJ 835, 54

\bibitem[Abdo A., et al. (2010)]{abdo2009} Torres, D. F., Cillis A. N., Mart\'in J. \& de O\~na Wilhelmi 2014, JHEAp, 1, 31
\bibitem[Abdo A., et al. (2010)]{abdo2009} Torres, D. F. 2018, Nature Astronomy in press, DOI: 10.1038 /
s41550-018-0384-5 (http://dx.doi.org); arXiv: astro-ph/1802.04177

\bibitem[Abdo A., et al. (2010)]{abdo2009} van den Bergh, S., 1978, ApJ, 220, 9
\bibitem[Abdo A., et al. (2010)]{abdo2009} Vigano, D, Torres, D. F., 2015, MNRAS 449, 3755
\bibitem[Abdo A., et al. (2010)]{abdo2009} Vigano, D, Torres, D. F., \& Martin, J. 2015, MNRAS, 453, 2599
\bibitem[Abdo A., et al. (2010)]{abdo2009} Vigano, D, Torres, D. F., Pessah, M., \& Hirotani K. 2015, MNRAS, 447, 1164

\bibitem[Abdo A., et al. (2010)]{abdo2009} Weiler, K. W. \& Seielstad, G. A., 1971, ApJ, 163, 455
\bibitem[Abdo A., et al. (2010)]{abdo2009} Weiler, K. W., \& Panagia, N. 1978, A\&A, 70, 419



\end{thebibliography}
\end{document}